\documentstyle[sprocl,psfig]{article}

\newcommand{\ber}{\begin{eqnarray}}
\newcommand{\eer}{\end{eqnarray}}
\newcommand{\grpp}{g_{\rho \pi \pi}}
\newcommand{\grdd}{g_{\rho DD}}
\newcommand{\gpdd}{g_{\pi DD^*}}
\newcommand{\gpnn}{g_{\pi NN}}
\newcommand{\grnn}{g_{\rho NN}}

\def\Journal#1#2#3#4{{#1} {\bf #2}, #3 (#4)}
\def\NPA{{\em Nucl. Phys.} A}
\def\PLB{{\em Phys. Lett.}  B}
\def\PRC{{\em Phys. Rev.} C}

\begin{document}

\title{
HADRONIC SCATTERINGS OF CHARM MESONS AND 
ENHANCEMENT OF INTERMEDIATE MASS DILEPTONS
}

\author{
ZIWEI LIN, CHE MING KO, BIN ZHANG
}

\address{
Cyclotron Institute and Physics Department, \\
Texas A\&M University, \\
College Station, TX 77843-3366, USA
} 


\maketitle\abstracts{
The scattering effects of charm mesons by hadrons such as the pion, rho meson
and nucleon are studied in an effective Lagrangian, and are found to be 
important in heavy ion collisions.
At the CERN-SPS energies, hadronic rescatterings are shown to harden 
the charm meson $m_T$ spectra, leading to a significant enhancement 
of the yield of intermediate-mass dimuons from charm meson decays.   
}

\section{Introduction}
Heavy quark production in hadronic reactions is 
reasonably well described by perturbative QCD. 
However, in heavy ion collisions, final-state interactions may affect
the spectra of produced heavy mesons. 
At the Relativistic Heavy Ion Collider 
(RHIC), a dense partonic system, often called the quark gluon plasma (QGP), 
is expected to be formed at the early stage.  
Since the QGP may induce a strong radiative energy loss of the produced 
heavy quarks, 
a change in the spectra of heavy meson observables 
could provide us information on the properties of the QGP.  
But interactions between heavy mesons and other hadrons 
may not be negligible and need to be studied.  

Such a hadronic modification of the charm spectra in heavy ion collisions
has been recently suggested~\cite{imr} 
as a possible explanation for the observed
enhancement of dimuons of intermediate masses in 
the NA50 experiments at the CERN-SPS.  
Assuming that charm mesons develop a transverse flow due to rescatterings
with hadrons, leading to a harder charm meson $m_\perp$ spectra, 
dimuons from charm meson decays are also found to have a harder 
$p_\perp$ spectrum. Based on the energy cuts for muons at the NA50 experiment, 
more dimuons would then be found to have an invariant mass above $1.5$ GeV.  
Another explanation based on dilepton productions from 
secondary meson-meson interactions
has also been proposed~\cite{gale}. 
Whether or not charm mesons acquire a transverse flow
depends on how strongly charm mesons interact with other hadrons 
during their propagation through the matter.  
In this study, we shall evaluate the scattering cross sections
of charm mesons with pion, rho, and nucleon, 
using an effective Lagrangian.     
The effects of hadronic scatterings on the 
charm meson transverse momentum spectra and dileptons from charm 
meson decays are then estimated for heavy ion collisions 
at CERN-SPS energies.

\section{Charm meson interactions with hadrons}

We consider the scattering of charm mesons ($D^+$, $D^-$, $D^0$, 
$\bar D^0$, $D^{*+}$, $D^{*-}$, $D^{*0}$, and $\bar D^{*0}$) 
with pion, rho, and nucleon. 
If the $SU(4)$ symmetry were exact, 
interactions involving pseudo-scalar and vector mesons 
could be described by the following Lagrangian:
\ber
{\cal L}_{PPV}=ig Tr \left ( P^\dagger V^{\mu \dagger} \partial_\mu P \right )
+ h.c. \;
\label{lsu4}
\eer
where $P$ and $V$ represent, respectively, the following $4\times4$ 
pseudo-scalar and vector meson matrices:
\ber
P&=&\left (
\begin{array}{cccc}
\frac{\pi^0}{\sqrt 2}+\frac{\eta}{\sqrt 6}+\frac{\eta_c}{\sqrt 6}
& \pi^+ & K^+ & \bar D^0 \\
\pi^- & -\frac{\pi^0}{\sqrt 2}+\frac{\eta}{\sqrt 6}+\frac{\eta_c}{\sqrt 6}
& K^0 & D^- \\
K^- & \bar K^0 & -\eta \sqrt {\frac{2}{3}}+\frac{\eta_c}{\sqrt 6}
& D_s^- \\
D^0 & D^+ & D_s^+ & -\frac{3\eta_c}{\sqrt 6} 
\end{array}
\right ) \;, \nonumber \\[2ex]
V&=&\left (
\begin{array}{cccc}
\frac{\rho^0}{\sqrt 2}+\frac{\omega}{\sqrt 6}+\frac{J/\psi}{\sqrt 6}
& \rho^+ & K^{*+} & \bar D^{*0} \\
\rho^- & -\frac{\rho^0}{\sqrt 2}+\frac{\omega}{\sqrt 6}+\frac{J/\psi}{\sqrt 6}
& K^{*0} & D^{*-} \\
K^{*-} & \bar K^{*0} & -\omega \sqrt {\frac{2}{3}}+\frac{J/\psi}{\sqrt 6}
& D_s^{*-} \\
D^{*0} & D^{*+} & D_s^{*+} & -\frac{3J/\psi}{\sqrt 6}
\end{array}
\right ) \;. \nonumber
\eer

Expanding the Lagrangian in Eq.(\ref{lsu4}) in terms of the meson fields
explicitly, we obtain the following Lagragians for meson-meson interactions:
\ber
&&{\cal L}_{\pi DD^*}=-i\gpdd \bar D^{* \mu} \vec \tau \cdot
\left [ D ( \partial_\mu \vec \pi) - ( \partial_\mu D) \vec \pi \right ]
+ h.c.\; , \nonumber \\
&&{\cal L}_{\rho DD}=-i\grdd
\left [ \bar D \vec \tau (\partial_\mu D) -(\partial_\mu \bar D) \right . 
\left . \vec \tau D \right ] \cdot \vec \rho^\mu \; ,\nonumber \\
&&{\cal L}_{\rho \pi\pi}=\grpp \vec \rho^\mu \cdot 
\left ( \vec \pi \times \partial_\mu \vec \pi \right ) \;. \nonumber
\eer
We also need the following Lagrangians for meson-baryon interactions: 
\ber
&&{\cal L}_{\pi NN}=-i\gpnn \bar N \gamma_5 \vec \tau N \cdot \vec \pi\; ,
\nonumber \\
&&{\cal L}_{\rho NN}=\grnn \bar N 
\left ( \gamma^\mu \vec \tau \cdot \vec \rho_\mu + \frac {\kappa_\rho}{2m_N} \sigma^{\mu \nu} \vec \tau \cdot \partial_\mu \vec \rho_\nu \right ) N
\; . \nonumber
\eer

\begin{figure}[ht]
\psfig{figure=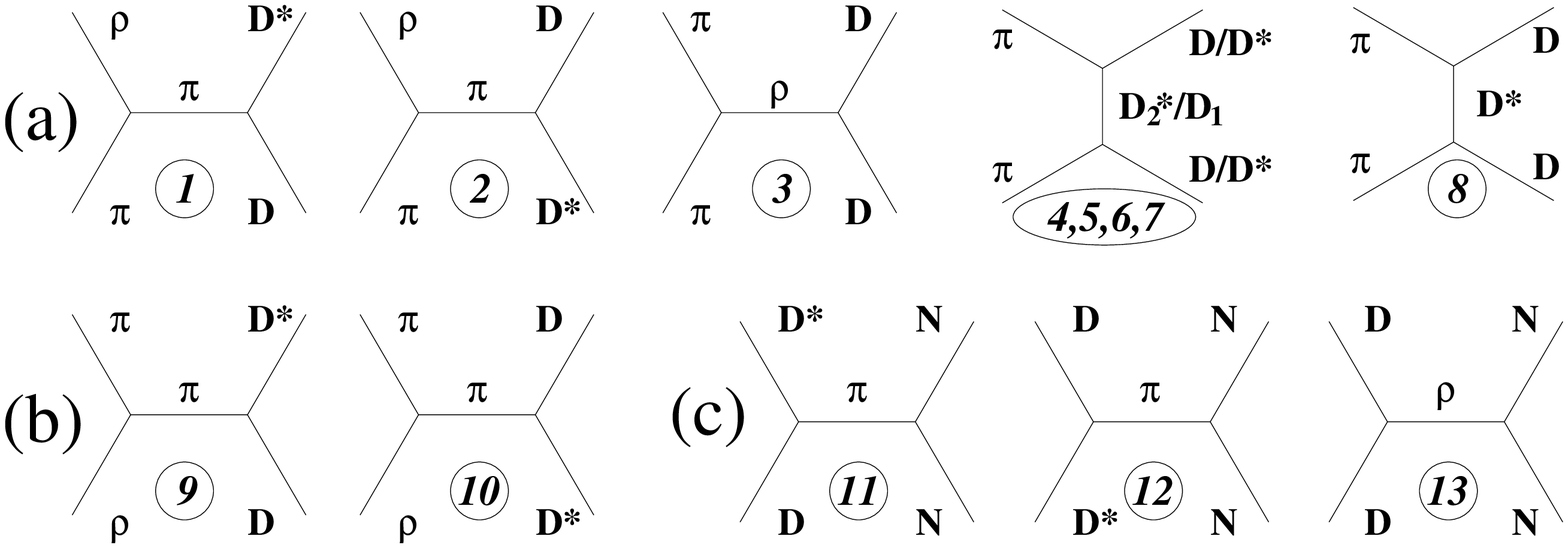,height=0.99in,width=4.65in}
\caption{ The included diagrams for (a) $D \pi$, 
(b) $D \rho$, and (c) $D N$ scatterings.
}
\label{diagrams}
\end{figure}

Fig.\ref{diagrams} shows the Feynman diagrams considered in this study for 
charm meson interactions with the pion (diagrams 1 to 8), 
the rho meson (diagrams 9 and 10), and the nucleon (diagrams 11 to 13).  
The differential cross sections of these processes can be found 
in the reference~\cite{ds}.  Possible interferences among diagrams 3, 4 and 8 
have not been included.

For coupling constants, we take $\grpp=6.1$, 
$\gpdd=4.4$, $\grdd=2.8$~\cite{rdd}, $\gpnn=13.5$, $\grnn=3.25$, 
and $\kappa_\rho=6.1$.  
The $SU(4)$ symmetry assumed in the Lagrangian in Eq.(\ref{lsu4})  
would give the following relations:
\ber 
g_{\pi KK^*}\! (3.3)\!=\!\gpdd \!(4.4)\!=\!g_{\rho KK} \!(3.0)
\!=\!\grdd \!(2.8)
\!=\!\frac  {g_{\phi KK}}{\sqrt 2}\! (3.4)\!=\!\frac {\grpp}{2} \!(3.0).
\nonumber
\eer
The empirical values given in parentheses agree reasonably well 
with this prediction even though the $SU(4)$ symmetry is badly broken.
Form factors are introduced at the vertices 
to take into account the structure of hadrons.
For $t$-channel vertices, monopole form factors of the form 
$f(t)=(\Lambda^2-m_\alpha^2)/(\Lambda^2-t)$ are used, 
where $\Lambda$ is a cut-off parameter~\cite{ds}, 
and $m_\alpha$ is the mass of the exchanged meson.
It should be noted that the cross sections for diagrams
2 and 9 ($D^* \pi \leftrightarrow D \rho$) are singular 
because the intermediate mesons can be on-shell.  
In the present study, we simply add an imaginary part of $50$ MeV 
to the mass of the intermediate pion as the regulator.  

\begin{figure}[ht]
\psfig{figure=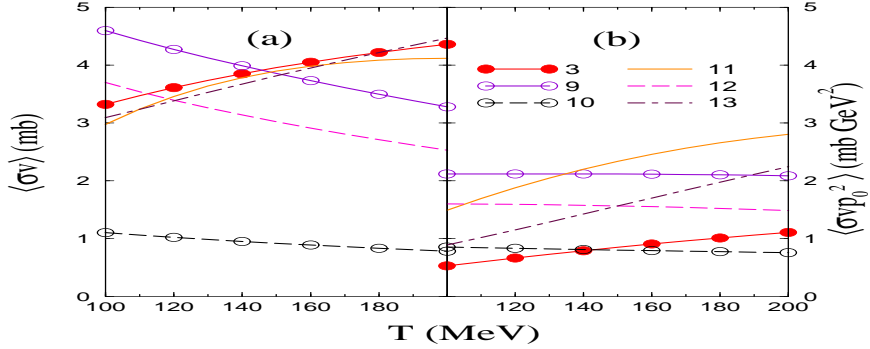,height=1.83in,width=4.6in}
\caption{
Thermal averages (a) $<\!\sigma v\!>$, and 
(b) $<\!\sigma v p_0^2\!>$ 
for the dominant charm meson scattering processes as a function of temperature.
The numbers in the notation of the curves correspond to the diagram numbers  
in Figure~\ref{diagrams}.
}
\label{sv}
\end{figure}

The thermal averaged cross section, $<\!\sigma v\!>$, 
is shown in Fig.\ref{sv}(a) for initial particles 
with a thermal distribution at temperature $T$.  
Only the dominant scattering channels which have values above $1.1$ mb  
are shown. 

\section{Estimates of Rescattering Effects}

In this section, 
we estimate the effects of hadronic rescatterings 
on both the charm meson $m_\perp$ spectra and 
the invariant mass distribution of dileptons from charm meson decays 
in heavy ion collisions at SPS energies.

We first determine the squared momentum transfer to a charm meson, $p_0^2$,
as the squared momentum of the final charm meson $D_2$ in the rest frame 
of $D_1$ for a scattering process $D_1 X_1 \rightarrow D_2 X_2$.   
In the charm meson local frame, we assume the time evolution 
of the hadron densities as $\rho(\tau) \propto 1/\tau$.
Then the total number of scatterings for a charm meson is given by
\ber
N = \int_{\tau_0}^{\tau_F} \sigma v \rho d\tau 
\simeq \sigma v \rho_0 \tau_0 
\ln \left ( \frac {R_A m^D_\perp}{\tau_0 p^D_\perp} \right ) \;,  
\nonumber
\eer
and the squared total momentum transfer from hadronic scatterings is
\ber
<p_{\rm S}^2>=<N p_0^2>=
\left[ \sum_{i=\pi,\rho,N\cdots}<\sigma v p_0^2>_i\rho_{i0}\right ]
\tau_0 \ln 
\left ( \frac {R_A <\!m^D_\perp\!>}{\tau_0 <\!p^D_\perp\!>} \right ) \;.
\nonumber
\eer
Thus, the relevant quantity is the thermal average 
$<\!\sigma v p_0^2\!>$ instead of 
the usual $<\!\sigma v\!>$.  
Fig.\ref{sv}(b) shows this thermal average for the dominant 
scattering channels which have values above $0.75$ mb$\cdot$GeV$^2$.  
Summing up contributions from all scattering channels 
in Fig.\ref{diagrams} (a), (b) and (c) separately, and simply 
dividing by 2 to account for the average over $D$ and $D^*$, we get 
\ber 
<\sigma v \;p_0^2> \simeq 1.1, 1.5 {\rm \;and \;} 2.7
{\rm \;mb \cdot GeV^2} \; {\rm at}\; T=150 \;{\rm MeV} 
\nonumber
\eer 
for $\pi$, $\rho$ and $N$ scatterings with charm mesons, respectively.    

For central $Pb+Pb$ collisions at SPS energies, 
the initial total numbers of particles are about
$500(\pi)$, $220(\rho)$, $100(\omega)$, $80(\eta)$, $180(N)$, $60(\Delta)$, 
and $130$(higher baryon resonances)~\cite{gq}. 
We have not calculated the scattering cross sections 
between charm mesons and hadrons such as kaons, $\omega$, $\eta$, $\Delta$,
and higher baryon resonances.     
For a conservative estimate of the effect, 
we only include $\pi$, $\rho$ and nucleon, and we obtain 
\ber
&&\rho_0 \tau_0 
\simeq 0.79(\pi), 0.35(\rho), {\rm \;and \;} 0.28({\rm nucleon}) 
{\rm \;fm^{-2}} \;, \nonumber \\
&&\Rightarrow <p_{\rm S}^2> \simeq 0.61 \;{\rm GeV^2}
\; \Rightarrow \; T_{\rm S} = 96 \;{\rm MeV}\; . \nonumber
\eer
In the above, we have taken $\tau_0=1$ fm. 
The parameter $T_{\rm S}$ characterizes the scattering strength  
and is given by $T_{\rm S} \simeq <\!p_{\rm S}^2\!>\!\!/(3m_D)$ 
in the lowest-order approximation~\cite{ds}.
Based on Monte Carlo simulations, $T_{\rm S}$ has been related to the inverse 
slope $T_{\rm eff}$ of the final charm meson $m_\perp$ spectrum, and this 
is shown in Fig.6 of Ref.~\cite{imr}.  
From that figure, we find that 
the charm meson $T_{\rm eff}$ increases from $160$~MeV to about $235$~MeV, 
leading to a dimuon enhancement factor of about $2.1$ 
for the NA50 acceptance.  

For heavy ion collisions at RHIC energies, 
in addition to hadronic rescatterings of charm mesons, 
partonic rescattering effects on charm quarks also need to be included.  
Furthermore, radiative processes of charm quarks inside the QGP 
would further complicate the issue as they may cause energy loss 
and soften the charm meson $m_\perp$ spectra.  
More studies are thus needed before one can make predictions for RHIC.

\section{Discussions and Summary}

From our calculated scattering cross sections of 
charm mesons with hadrons such as pion, rho meson and nucleon, 
we have given an estimate of the rescattering effect 
at the SPS energies.     
We find that hadronic rescatterings in heavy ion collisions 
are likely to have a significant effect on charm meson spectra, 
and also the dilepton spectra from charm meson decays.

The estimates given above are, however, based on a simple assumption 
on the time evolution of the hadronic system,   
which enables us to make a more analytical estimate 
for the rescattering effects.  
For a quantitative study of the rescattering effects on charm 
meson observables, 
studies with a partonic and hadronic cascade program
are much needed 
as the time evolution and the chemical equilibration 
of the dense system 
are better simulated in such~a~model. 
 
\section*{Acknowledgments}
This work was supported by the National Science 
Foundation under Grant No. PHY-9870038, the Welch Foundation under Grant
No. A-1358, and the Texas Advanced Research Project FY97-010366-068.

\end{document}